# Standoff chemical identification of harmful aerosols

Garrett C. Mathews[1*], Alyssa Lalko[1*], Anna C. Ziola[2*], Masayuki Takeuchi[1*], Leo C. Bloxham[2], Douglas A. Day[2], Anne V. Handschy[2], Amanda Makowiecki[3], Kevin Williamson[3], Nazanin Hoghooghi[1,4], Scott Egbert[1], Peter Chang[5], Tsung-Han Wu[5], Anthony Harness[3], Robert Wright[3], Jiang Li[6], Tin Komljenovic[7], John Bowers[8], Satoshi Takahama[9], Daven Henze[1], Kerry Vahala[10], Scott A. Diddams[5], Jose L. Jimenez[2], Gregory B. Rieker[1,3]

[*]These authors contributed equally to this work
[1]Paul M. Rady Department of Mechanical Engineering, University of Colorado Boulder, Boulder, CO, USA
[2]Department of Chemistry and Cooperative Institute for Research in Environmental Sciences (CIRES), University of Colorado Boulder, Boulder, CO, USA
[3]Longpath Technologies Inc., Boulder, CO, USA
[4]Time and Frequency Division, National Institute of Standards and Technology, Boulder, CO, USA
[5]Department of Electrical, Computer, and Energy Engineering, University of Colorado Boulder, Boulder, CO, USA
[6]hQphotonics Inc., Arcadia, CA, USA
[7]Nexus Photonics, Goleta, CA, USA
[8]Department of Electrical and Computer Engineering, University of California Santa Barbara, Santa Barbara, CA, USA
[9]Laboratory for Atmospheric Processes and Their Impacts, École Polytechnique Fédérale de Lausanne, Lausanne, Switzerland
[10]T. J. Watson Laboratory of Applied Physics, California Institute of Technology, Pasadena, CA, USA

**Aerosols – solid or liquid particles suspended in air – are an effective means of dispersing chemical warfare agents and toxic pharmaceuticals, posing acute risks to public safety and security[1–5]. Exposure to aerosol particulate matter in the environment and in industrial settings has also been linked to respiratory and cardiovascular disease[6–9]. In both cases, rapid chemical identification is key to generating early warnings and attributing sources. Established methods chemically identify aerosols only after extracting a sample, while standoff techniques such as lidar classify aerosols only into broad categories. Chemical identification at a distance is hindered by aerosol extinction signatures that are weak, spectrally broad, and obscured by far stronger absorption from ambient gases. Here, we demonstrate time-resolved, standoff chemical differentiation of aerosols with dual-comb spectroscopy and machine learning identification algorithms. The high spectral resolution of dual-comb spectroscopy allows ambient gas absorption to be removed, and the identification algorithms enable chemical-specific identification without knowledge of the aerosol properties that affect extinction signatures (size and morphology). We demonstrate identification of aerosolized benzyl fentanyl (a chemical simulant of the dangerous synthetic opioid fentanyl) in mixtures with other aerosolized compounds in a standoff measurement through ambient $H_2O$, $CO_2$, and $CH_4$. Despite the aerosol producing only subtle, <1% changes in the extinction spectrum across hundreds of wavenumbers, we find the limit of identification to be an order of magnitude below lethal concentrations for a 1 m aerosol**

**cloud. This technique provides a new paradigm for performing rapid chemical speciation of aerosols using their infrared absorption signatures without requiring extractive sampling. Together with advances in chip-based mid-infrared dual-comb spectrometry, we anticipate this technique will enable real-time networks for standoff detection of aerosol emissions and aerosolized chemical threats.**

Chemical discrimination of aerosols is necessary to alert individual or population exposure to toxic substances, assessment of health risks, and identification of potential sources. This capability is therefore critical for implementing timely measures such as public advisories, evacuations, and medical treatment. Chemical discrimination is typically performed through mass spectrometry or optical spectroscopy. Mass spectrometry is widely used to identify aerosol chemical constituents based on their mass-to-charge ratio following ionization[10–13]. Spectroscopic aerosol speciation is performed using Fourier transform infrared spectroscopy (FTIR), which measures the absorption of infrared light at wavelengths corresponding to fundamental vibrational frequencies[14–17], or Raman spectroscopy, which measures the wavelength shift of scattered photons induced by interactions with molecular vibrations[18–21]. However, these techniques require extracting an aerosol sample from the native environment to perform ionization for mass spectrometry, eliminate ambient gas interference for FTIR, or generate adequate Raman scattering signals. Extractive sampling is not amenable to rapid, continuous monitoring over large spatial scales, such as entire manufacturing plants or large population centers. Sample extraction can also have negative effects on the sample itself, such as perturbing the chemical composition.

Standoff aerosol sensors based on Light Detection and Ranging (LIDAR)[22–26] and Open-Path FTIR (OP-FTIR)[27–31] have been applied to detect aerosols, but their identification capabilities are limited to broad aerosol classifications (e.g., smoke, dust, ash, urban sources, maritime sources) rather than chemical-specific identification. Standoff aerosol chemical identification has been demonstrated using FTIR measurements of broadband laser spectra scattered from aerosol particles[32,33], but the standoff distance was limited to 10s of cm.

The key challenge for chemical differentiation of aerosols is the extremely weak and subtle extinction signatures that they impart on light, due to their relatively low concentrations and large molecular structure compared to ambient greenhouse gases (e.g., $H_2O$, $CO_2$, $CH_4$). *Standoff* infrared measurements must additionally overcome spectral interference from these small gases, which produce strong, spectrally narrow absorption signatures at atmospheric conditions. A satellite-based Fourier transform spectrometer (FTS), the Atmospheric Chemistry Experiment FTS (ACE-FTS), which measures sunlight passing edge-on through the upper atmosphere, has distinguished several classes of atmospheric particles by composition (ice, nitric acid trihydrate, sulfuric acid solutions, and smoke)[34]. However, this requires a physically large, high-resolution instrument that is not well-suited for routine field deployment due to the inverse relationship between spectral resolution and the scanning arm length of the FTS Michelson interferometer. Spectral interference from small gas-phase molecules can be mitigated for FTIR measurements in laboratory aerosol chambers by reducing the concentration of interfering gases[31,35], but this is again not practical for outdoor field deployments.

We address the need for a scalable standoff measurement technique capable of aerosol chemical identification by employing mid-infrared (MIR) dual-comb spectroscopy (DCS) and novel machine learning aerosol identification algorithms to identify the chemical species present in aerosol mixtures without prior knowledge of the sample or aerosol size and morphology properties. DCS is an emerging technique that utilizes optical frequency combs that emit thousands of discrete, coherent wavelengths of light to measure the spectral absorbance or extinction of a sample with high spectral resolution over a broad bandwidth[36–38]. DCS has recently been applied to open-path monitoring of greenhouse gases[39–42] and broadband multi-species mid-infrared measurements[43–46]. DCS has several inherent advantages for open-path aerosol sensing. The combination of high spectral resolution and bandwidth can resolve sharp absorption signatures from small gases from broad absorption signatures associated with aerosols. The coherent output of a frequency comb laser can be transmitted over long distances for large-area monitoring, and the high intensity of the coherent beam enables measurements of absorption or extinction without requiring a solution to the full radiative transfer equation. DCS measurements can be acquired quickly, due in part to the lack of moving parts, which enables monitoring of dynamic systems. In addition, frequency combs can be translated to integrated photonic platforms to achieve low size, weight, and power instruments well-suited for a distributed sensing network.

Here, we demonstrate standoff DCS measurements of aerosol extinction spectra in controlled chambers with natural air (containing $H_2O$, $CO_2$, and $CH_4$) and apply machine learning algorithms to identify aerosol species in neat aerosols and external mixtures. Benzyl fentanyl, a surrogate for the synthetic opioid fentanyl, is successfully identified as a component in the aerosol mixtures to demonstrate standoff detection of aerosolized pharmaceutical threats. Time-resolved measurements of benzyl fentanyl mixing in the aerosol chamber demonstrate the measurement of a dynamic event with one-second time resolution. This work presents the first time-resolved, chemical-specific standoff aerosol measurements, which is a novel approach for aerosol sensing with broad applicability in aerosol science, aerosol monitoring, and aerosol threat detection.

## Dual-Comb Spectroscopy for Aerosol Chemical Speciation

Figure 1 illustrates the operating principle of standoff DCS aerosol measurements and the procedure to identify aerosol chemical species from the measured spectra. Broadband laser spectra emitted from two MIR frequency combs are combined and transmitted across an open path containing atmospheric gases and aerosols from various sources. The light is reflected with a retroreflector and returns to the transceiver where it is collected and measured at high spectral resolution via heterodyne beating between pairs of comb teeth. Gases and aerosols present along the beam's line-of-sight impart unique absorption and extinction signatures on the laser spectrum. Greenhouse gases produce strong, spectrally narrow absorbance peaks due to their small molecular structure (which results in distinct absorption signatures) and their high concentrations (up to ~10 g/m³ for $H_2O$). Molecules with larger structures, including both aerosols and larger trace gases, will produce broad signatures, and their relatively low concentrations (~1–100 μg/m³) result in small peak absorbances. For this work, only aerosols are measured, but the same technique is applicable to other compounds which produce broad, weak absorption signatures such as organic gases.

To identify aerosol chemical species in the measured laser spectrum, the data is first processed to remove the spectrally narrow gas-phase absorption signatures and calculate the pure aerosol extinction spectrum. This is enabled by the high resolution of the DCS measurements, which fully resolve each individual gas-phase absorption transition without imparting a broad instrument lineshape. The pure aerosol extinction spectrum is then passed through a neural network trained with simulated aerosol spectra to identify specific chemicals present along the laser line-of-sight. Additional descriptions of DCS measurement fundamentals, data processing procedures, and identification algorithms are provided in the Methods section.

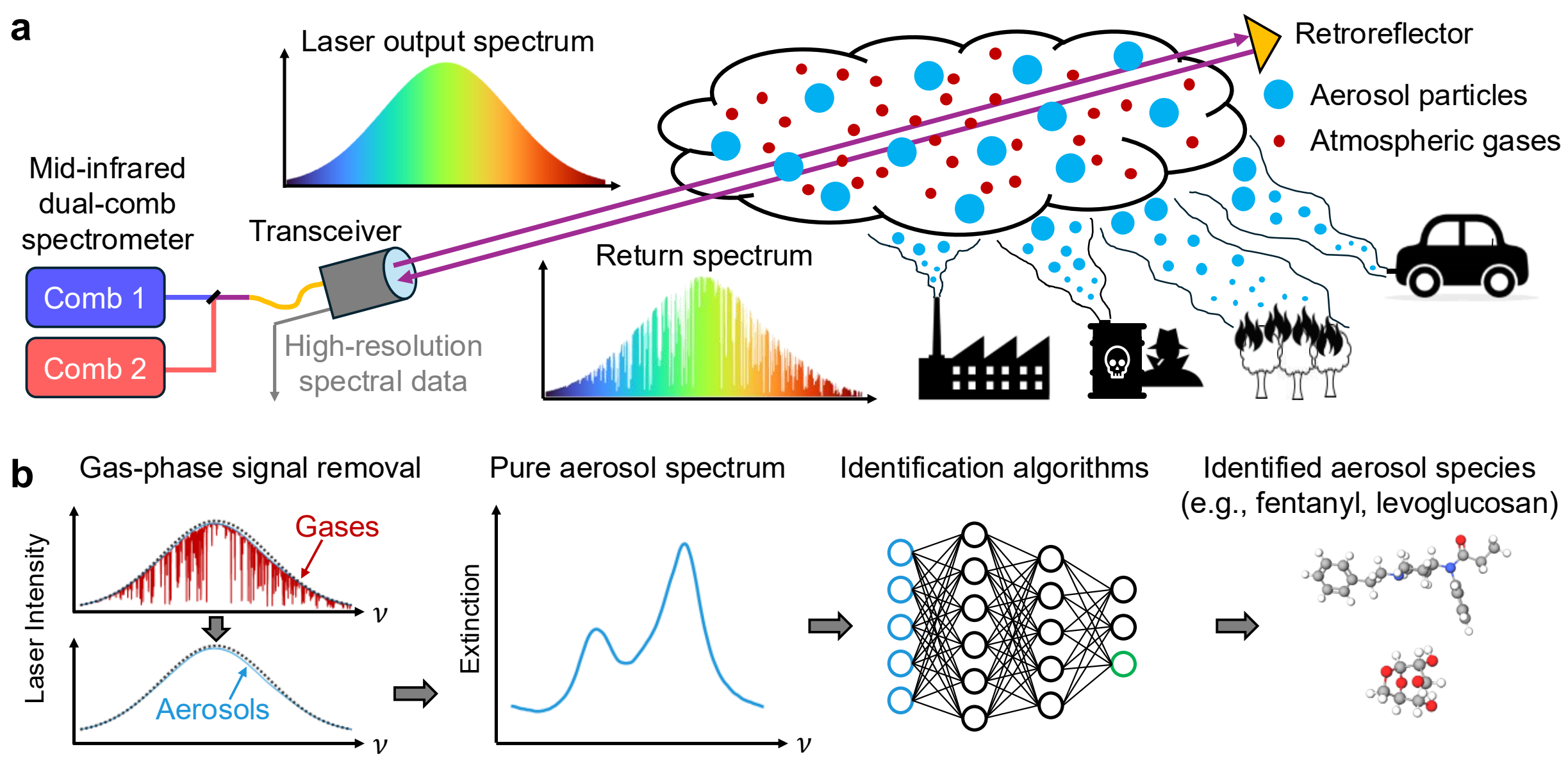

*Figure 1: (a) Operating principle of DCS for aerosol measurements. Two MIR frequency combs are combined and passed over an open path containing atmospheric gases and aerosols from a variety of sources. The light is then reflected back to the transceiver using a retroreflector, and the returned laser spectrum is measured. (b) Data processing procedure to perform chemical identification. Gas-phase signals are removed from the measured spectrum, enabling calculation of the pure aerosol extinction spectrum. The pure aerosol extinction spectrum is passed through a neural network to identify individual chemical species in the measurement.*

Standoff measurements of aerosols in an aerosol chamber were acquired using a custom dual-comb spectrometer to demonstrate the technique. Aerosols were generated, mixed, characterized and sustained within an enclosed, cylindrical, steel chamber. The aerosol chamber was 30 cm in diameter and 1.57 m long. The DCS laser light was directed through 5 cm diameter ZnSe windows installed on both sides, allowing it to pass through the chamber length, strike a retroreflector outside the chamber, then return through the chamber, giving an effective path length inside the chamber of 3.14 m. Aerosols were generated with different methods, depending on the compound, concentration, and aerosol size of interest, and were characterized for size and mass concentration in real time by an aerodynamic particle sizer (APS) and scanning mobility particle sizer (SMPS).

Two MIR frequency comb lasers produced spectra spanning 3.0–3.8 μm (2632–3333 $cm^{-1}$) with 21,300 comb teeth spaced at 0.033 $cm^{-1}$ (1 GHz)[47]. The output spectra were tuned to this spectral region to coincide with the C-H stretch vibrational mode, but operation has previously also been demonstrated at longer wavelengths spanning 7–13 μm[47]. The output of each frequency comb was combined into an optical fiber and passed to a custom transceiver designed to transmit light to the aerosol chamber and collect the back-reflected beam. The transceiver was placed at 2 m from the chamber, which was a physical limit of the laboratory, not a fundamental limit of the system or transceiver (100 m standoff is shown in the Methods section). The transceiver included two mercury cadmium telluride (MCT) detectors: one to measure a reference spectrum before and the other to measure the signal spectrum after transmission through the aerosol chamber. The reference spectrum measurement accounts for small changes in the laser output spectrum over time, which can generate subtle changes in transmission of similar magnitude and shape as the aerosol signatures of interest. For each measurement scenario, background data was acquired both before and after aerosol measurements. Measurement and background data were recorded continuously for at least five minutes for each acquisition.

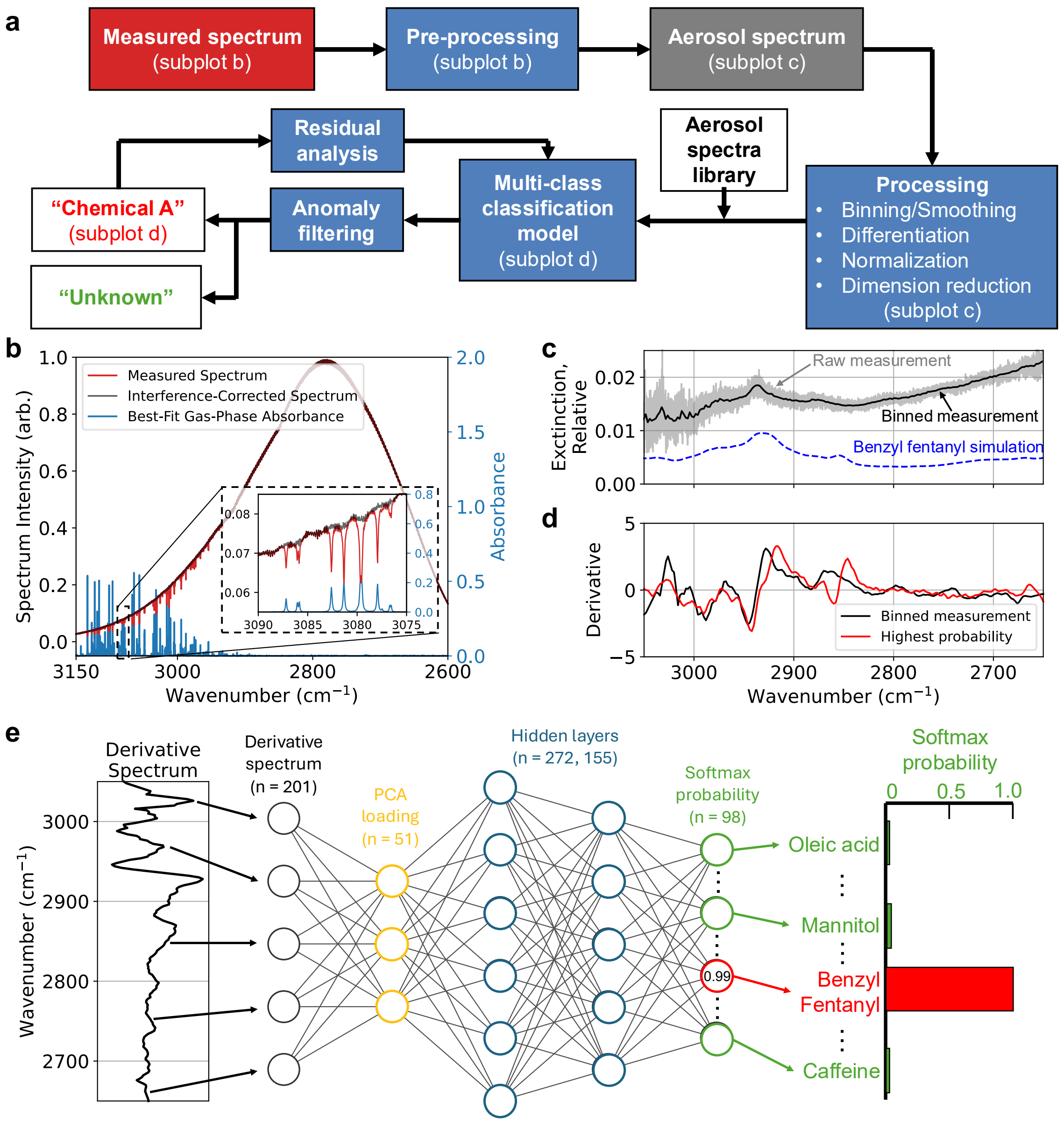

*Figure 2: (a) Flowchart illustrating the data processing workflow to identify chemicals present in measured spectra. (b) Measured transmitted intensity spectrum, best-fit gas-phase absorbance (from ambient species), and interference-corrected spectrum (with gas-phase absorption removed). (c) The relative extinction spectrum calculated from the interference-corrected transmission spectrum. Relative extinction was calculated by subtracting the extinction value measured at 3200 cm$^{-1}$ from the measured spectrum. A simulated benzyl fentanyl extinction spectrum is shown for reference. Details of the simulation are provided in the Methods section. (d) The first derivative spectrum and the highest-probability derivative spectrum returned by the neural network (benzyl fentanyl) are shown on the bottom plot in red. (e) Diagram of the neural network used to identify chemicals from the derivative spectrum.*

Figure 2 illustrates the data processing workflow used to identify aerosol compounds from measured spectra and shows example data acquired with benzyl fentanyl in the chamber at the different stages of data processing. Following the data processing flowchart shown in Figure 2(a), measured spectra were first passed through a pre-processing routine to remove absorption signatures from small gas-phase molecules and normalized by background data to eliminate interference and calculate a pure aerosol extinction spectrum. Identification of unknown aerosols from a measured extinction spectrum was then performed via a supervised machine learning algorithm. The algorithm consists of multiple steps: measured data curation to prepare input values for machine learning, event detection to determine whether or not the measured spectrum is background, multi-class classification to predict the identity of the chemical species, and anomaly filtering to avoid misclassification. For the training dataset, we prepared a library of extinction spectra of approximately 100 target chemicals, including benzyl fentanyl, with varying size distribution and mass concentration, based on either measured or estimated complex refractive index and Mie theory[48]. Within this algorithm, a critical challenge was accounting for aerosol effects on extinction spectra. While most work using infrared sensing for chemical characterization has focused on homogeneous samples[49–52], for aerosols, properties such as the particle size distribution and morphology greatly alter the shape of the extinction spectra[33,53]. To minimize the influence of these effects, we developed an approach using derivative spectra, as these are much more similar for a given chemical regardless of aerosol properties. Another hurdle was the reliance of the supervised machine learning algorithm on the set of chemicals present in the training database. This limitation was addressed by incorporation of an anomaly filtering step that prevented forced misclassification of unknown chemicals and provided greater confidence in the algorithm prediction. Lastly, to address multi-component mixtures, we expanded our chemical identification algorithm to allow prediction of multiple chemicals in aerosol mixtures via a hierarchical approach that sequentially identifies chemicals based on the presence of their most dominant spectral features.

Figure 2(b) and Figure 2(c) show measured data before and after data pre-processing. The measured spectrum in Figure 2(b) (plotted in red) shows clear signatures of gas-phase absorption from 2900 to 3200 $cm^{-1}$, but the aerosol signatures are not readily visible. A spectroscopic model including contributions from $H_2O$, $CO_2$, and $CH_4$ was fit to each measured spectrum, and the best-fit absorbance (plotted in blue) was removed from the measured intensity to correct for absorption signatures from small gases along the line-of-sight. The corrected intensity spectrum (plotted in black) illustrates the robustness of the gas-phase correction procedure. The $H_2O$ concentration measured using the DCS signal was approximately 3700 $mg/m^3$, which is two orders of magnitude larger than the benzyl fentanyl concentration of 38 $mg/m^3$ (measured with the APS and SMPS). Figure 2(c) shows the aerosol extinction spectrum calculated using the corrected spectra from a test and background measurement, and a simulated benzyl fentanyl extinction spectrum is shown for reference. The optical interference fringes (i.e., etalons) present in the corrected intensity spectra largely cancel out when the spectra are normalized to calculate the aerosol extinction spectrum, and remaining etalons are filtered out using the time-domain representation of the extinction spectrum[54]. After normalization and etalon removal, the spectrum is bin-averaged from a native spectral resolution of 0.033 $cm^{-1}$ to a resolution of 2 $cm^{-1}$. Since aerosol spectra are always broad and narrow gas-phase signals were removed, this averaging increased the signal-to-noise ratio (SNR) without altering the signals of interest. Measured extinction spectra are presented as relative extinction, calculated by subtracting the extinction value measured at 3200 $cm^{-1}$ (at the bin-averaged resolution), to remove non-physical negative extinction values that were occasionally produced during the pre-processing procedure. The apparent negative extinction values were due to changes in the returned laser power between the test and background datasets and can be accounted for with a DC or linear offset in the extinction spectrum. The final measured extinction spectrum exhibits clear peaks near 2935 and 2975 $cm^{-1}$ characteristic of benzyl fentanyl absorption. The signature is identified despite subtle, <1% changes in the extinction spectrum over hundreds of wavenumbers of optical spectrum.

Automated aerosol identification is performed by calculating the derivative spectrum from the bin-averaged extinction spectrum and passing the derivative spectrum to the input layer of the principal-component-coupled neural network. The derivative spectrum is shown in Figure 2(d). The neural network contains one principal component analysis (PCA) loading layer with 51 nodes, and two hidden layers with 272 and 155 nodes. Each node of the output layer corresponds to a likelihood that a given chemical component is present in the measured extinction spectrum. For the data shown, the highest likelihood output node corresponds to benzyl fentanyl with a softmax probability >0.99. All other output likelihoods are negligible, indicating a clear identification of benzyl fentanyl for this data.

In practical aerosol monitoring or threat detection, an aerosol of interest will be mixed with other aerosol compounds along the line-of-sight, and the ability to identify multiple aerosols in a mixture is critical. This capability was demonstrated by adding benzyl fentanyl to a test chamber already containing aerosolized oleic acid, a common organic aerosol produced from cooking sources[55], whose overlapping C-H vibrational signatures present a challenging spectral background. Oleic acid was initially added to the chamber, followed by a series of discrete additions of benzyl fentanyl, and a DCS measurement was acquired after each addition. The total aerosol mass concentration measured with the APS and SMPS is plotted in Figure 3(a). The aerosol column density, calculated as the product of aerosol concentration in $mg/m^3$ and the aerosol chamber length of 1.57 m, is also reported. The column density enables scaling for different aerosol cloud sizes because the extinction signatures scale linearly with both path length through the aerosol cloud and aerosol concentration at low optical depths. DCS measurements were acquired for five minutes after each benzyl fentanyl addition, and shaded regions in the concentration plot show four distinct time windows for data acquisition corresponding to the spectra plotted in Figure 3(b).

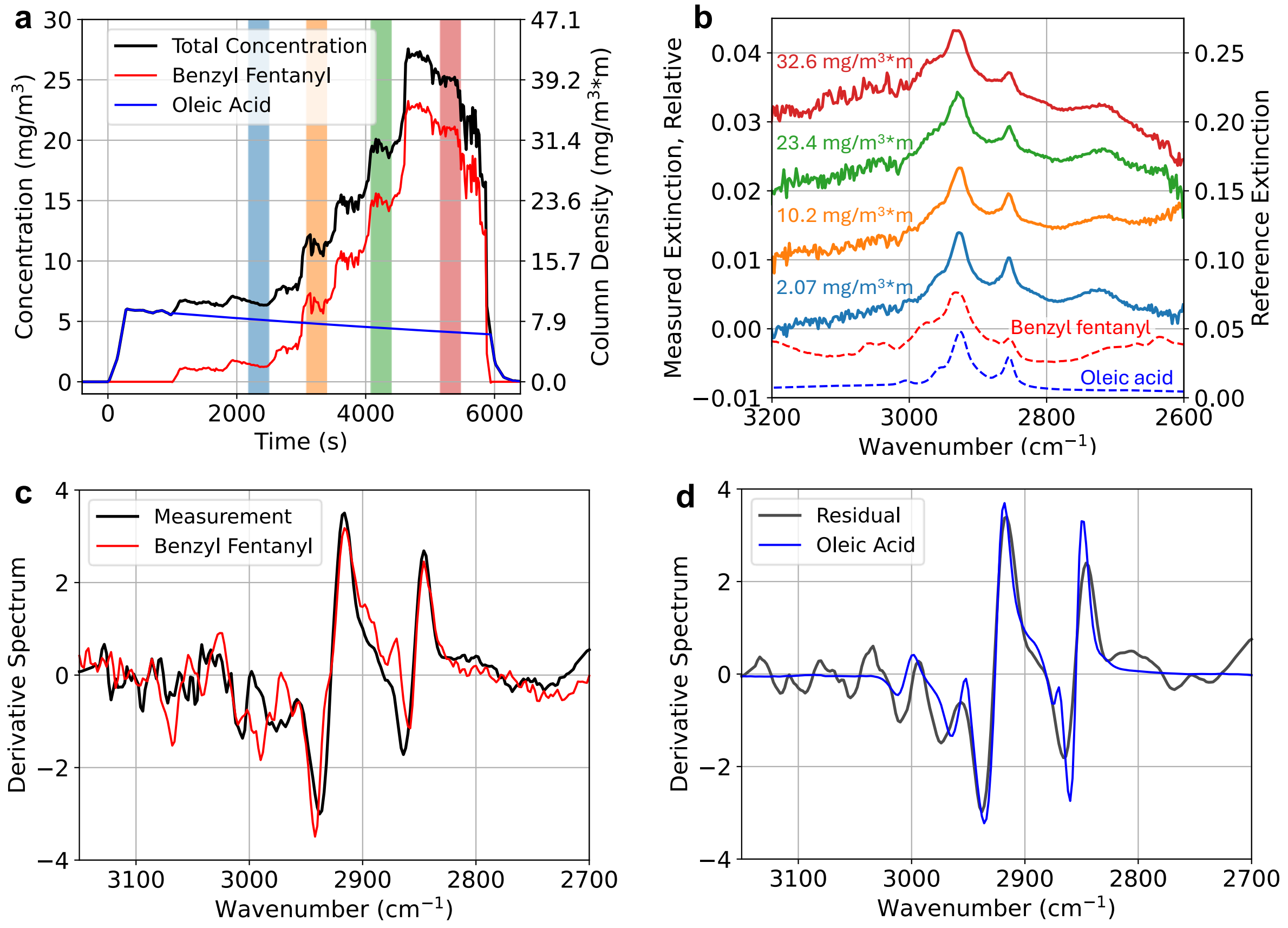

*Figure 3: Identification of benzyl fentanyl in a mixture with oleic acid. (a) Aerosol mass concentration measured with SMPS and APS instruments sampling from the aerosol chamber and the calculated aerosol column density. (b) Aerosol extinction spectra measured with DCS and reference absorbance spectra of pure oleic acid and pure benzyl fentanyl showing similar C-H stretch vibrational overlap. Measured extinction spectra are shown relative to the extinction measured at 3200 $cm^{-1}$, and each subsequent plot is shifted up by 0.01 in absorbance units for clarity. (c) Derivative spectrum and identification of benzyl fentanyl using data acquired from 3090–3300 seconds (orange band) after the start of the test, corresponding to a benzyl fentanyl column density of 10.2 $mg/m^3$*m. (d) Derivative of residual spectrum after subtraction of benzyl fentanyl and identification of oleic acid.*

Figure 3(b) shows the extinction spectra calculated from data acquired during each time window indicated in Figure 3(a), and the column density of benzyl fentanyl is listed for each measured spectrum. Simulated extinction spectra of pure oleic acid and pure benzyl fentanyl are shown for reference and demonstrate the overlapping C-H stretch vibrational features of the two aerosols. Measurements acquired with low benzyl fentanyl column densities are dominated by the spectral signatures of oleic acid as indicated by the strong peaks near 2850 and 2915 $cm^{-1}$. As the benzyl fentanyl loading is increased, the measured spectrum exhibits signatures corresponding to the benzyl fentanyl in addition to the oleic acid as indicated by the appearance of additional spectral features near 2975 and 3015 $cm^{-1}$. Extinction spectra measured after each benzyl fentanyl addition were passed to the identification algorithm to determine the limit of identification (LOI; lower bound to determine the identity of chemicals with no prior knowledge of chemical composition) of benzyl fentanyl in the mixture. A hierarchical identification approach was used to identify each component in the mixture following the procedure outlined in Figure 2(a). After the identification of a single aerosol species, the residual spectrum (i.e., measured spectrum minus the spectrum of the identified species via asymmetric least squares approach) was passed through the neural network to identify additional aerosol species. Figure 3(c) shows the derivative spectrum first passed through the identification neural network and the simulated derivative spectrum of benzyl fentanyl, which was identified in the first pass. Figure 3(d) shows the derivative of the spectrum after removing the benzyl fentanyl component and the simulated derivative spectrum of oleic acid, which was identified on the second pass. Despite the similar structures in the derivative spectra, the identification algorithm was able to identify each component individually using the hierarchical approach. Following this procedure, benzyl fentanyl was first identified in the measurement with a benzyl fentanyl column density of 10.2 mg/m$^3$*m (5 min measurement time), which defines the LOI in this mixture.

To demonstrate the capability of monitoring transient physical or chemical processes, DCS measurements were acquired with a time resolution of one second while benzyl fentanyl was added to the aerosol chamber and mixed with an internal fan. For each measured spectrum, the benzyl fentanyl absorbance magnitude was determined by fitting a reference spectrum to the measured data. The reference spectrum was measured using attenuated total reflectance (ATR) FTIR, and the magnitude of the reference spectrum was scaled following a nonlinear fitting algorithm to find the best fit to the measured data. Figure 4(a) shows a time history of the benzyl fentanyl absorbance magnitude, which scales linearly with the benzyl fentanyl concentration, normalized to a magnitude of one at steady state. The absorbance magnitude is initially centered around zero before and during the addition of benzyl fentanyl. The absorbance magnitude then shows a sharp increase over approximately 20 seconds as the aerosolized benzyl fentanyl particles are moved into the laser beam path via the mixing fan, and the magnitude remains steady after complete mixing. The steady-state benzyl fentanyl concentration was 44 mg/m$^3$, measured using the APS and SMPS. Figure 4(b) shows an example of an extinction spectrum measured with a time resolution of one second and the corresponding best-fit spectrum. Benzyl fentanyl absorption features are difficult to identify in the native-resolution spectrum, but they emerge clearly near 2940 and 2980 cm$^{-1}$ after applying bin averaging. The high spectral resolution of the DCS measurement is helpful here because it enables high resolution correction for sharp gas-phase interference. Then, bin averaging across optical frequencies identifies the subtle and broad absorption signatures of the aerosol at much higher SNR than at the native measurement resolution. The limit of detection (LOD) for benzyl fentanyl with this data analysis method is 40 mg/m$^3$*m (in 1 s), defined as three times the standard deviation of the absorbance signal magnitude measured before mixing multiplied by the steady-state benzyl fentanyl concentration and the chamber path length. The higher LOD here compared to the LOI for the mixture data is primarily due to the averaging time being reduced by a factor of 300. If the LOD is scaled to account for this difference, the LOD is estimated as 2.3 mg/m$^3$*m (in 5 min). The lower 5-minute LOD compared to the mixture LOI stated before may be attributed to the presence of only a pure compound here.

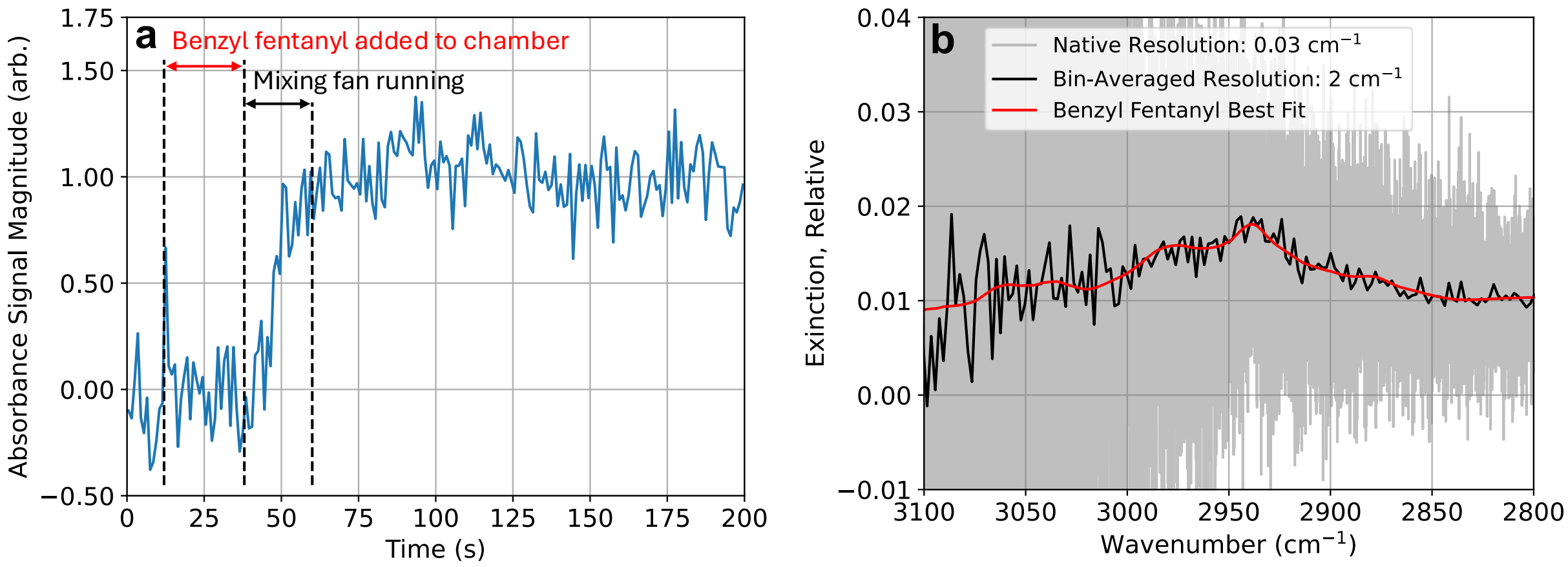

*Figure 4: (a) Time history of the benzyl fentanyl absorbance magnitude measured while benzyl fentanyl was added to the aerosol chamber and the mixing fan inside the chamber was running. (b) An example extinction spectrum measured with 1-second time resolution and the resulting best-fit benzyl fentanyl spectrum.*

## Discussion

DCS provides high-resolution measurements of aerosol extinction spectra in a standoff configuration through atmospheric gases, and the combination of this data with novel machine learning algorithms enables time-resolved chemical-specific aerosol identification. High-quality measurements of aerosol extinction spectra are enabled by the high spectral resolution and simultaneously broad spectral output that resolve strong gas-phase interference and subtle aerosol absorption bands at the same time. The high spectral resolution is key to removing the narrow gas-phase absorption signals to produce an isolated aerosol extinction spectrum. In instruments with lower spectral resolution, narrow gas-phase absorption signatures are mixed with the aerosol extinction because the measured spectrum is convolved with the instrument lineshape function. In contrast, DCS samples the spectrum with dense point spacing (~0.01 $cm^{-1}$), and each sampled point has a negligible instrument lineshape (~$1\times10^{-6}$ $cm^{-1}$). Measurements thereby fully resolve the narrow absorption signatures, which can then be removed after fitting high-fidelity absorption models to the measured spectra. As shown in Figure 2, narrow gas-phase signatures are several orders of magnitude larger than the aerosol signatures of interest and can be effectively removed from the measured spectra. Furthermore, the removal of narrow gas-phase absorption signatures enables averaging across optical frequencies, leveraging the high spectral resolution of DCS to reduce the noise in the pure aerosol spectrum, which only contains subtle, spectrally broad signatures.

The aerosol identification neural network leverages the low-noise aerosol extinction measurement to identify the aerosol chemical species present in the sample without knowledge of the aerosol size or morphology. This work focuses on chemical identification rather than quantification. However, quantitative gas measurements are commonly performed with DCS[39,40], and further refinement of this system could also enable quantitative aerosol concentration, size and morphology measurements.

The high time resolution of DCS measurements provides a unique benefit in situations where time dynamics are of interest, particularly if target species are known *a priori*, because the LOD can be lower than the LOI. High-time-resolution measurements can be applied to aerosol monitoring where a specific chemical species, or set of chemical species, is expected to be present, such as toxic manufacturing byproducts where fast alerts are required, or emission sampling near sources where concentrations rapidly shift with wind direction[56]. In laboratory measurements, this can be applied to study the time dynamics of aerosol chemical reactions where the reactants, intermediates, and product species are known. DCS presents an advantage over FTIR-based measurements where high resolution requires acquisition times on the order of minutes[57] and an advantage over sampling-based methods in cases where short-lived reactive intermediate species cannot be sampled.

This DCS system is currently well-suited for monitoring aerosol chemical threats given the experimentally determined LOI on the order of 10 $mg/m^3$*m (in 5 min), validated with measurements of benzyl fentanyl in an aerosol mixture. The estimated lethal dose of fentanyl is 2 mg. At a nominal breathing rate of 6 L/min, a concentration of 667 $mg/m^3$ would be required to deliver this dose in one minute (assuming 50% deposition[58]). This is comparable to several other chemical warfare agents (e.g., VX, sarin, sulfur mustard) which have lethal respiratory concentrations on the order of 10s to 1000s of $mg/m^3$*min [2]. Even given a relatively small, uniform aerosol cloud length on the order of 1 m, the current LOI for this system is an order of magnitude lower than the lethal mass concentration.

In the context of air quality monitoring, this DCS system enables chemical differentiation of C-H-containing compounds. This capability was demonstrated by the measurement of the aliphatic C-H band (near 2800–3000 $cm^{-1}$) and the weaker aromatic C-H band (near 3050 $cm^{-1}$) of benzyl fentanyl and the chemical differentiation from aerosolized oleic acid. This ability enables monitoring and identification of C-H compounds in particulate matter (PM) and classes of hazardous air pollutants (HAPs) partitioning to the condensed phase, such as polycyclic aromatic hydrocarbons and pesticide aerosols. Extending the frequency comb output to longer wavelengths[47] would enable chemical differentiation using additional functional groups. To demonstrate outdoor air quality monitoring over extended open paths, a measurement of aerosolized dibutyl phthalate, classified as a HAP, was performed using a second, fieldable MIR DCS system at a standoff distance of 100 m. Additional details and results from this measurement are provided in the Methods section.

Mid-infrared frequency comb sources are new and rapidly advancing[46,47,59,60], and lower limits of identification can be achieved by improving the SNR of the DCS system. Techniques leveraging higher output power MIR frequency comb sources[46], using improved detection approaches[61] and employing frequency comb operation modes that enable faster acquisition and averaging of DCS data[62] can all potentially improve the measurement SNR. Finally, the laser-based nature of DCS is also amenable to cavity enhancement[63], which can significantly increase the interaction path length and, therefore, reduce the aerosol concentration LOI. While this would require a sampling configuration, rather than standoff, cavity-enhanced DCS aerosol measurements could potentially compete with conventional mass spectrometer sensitivity while improving time resolution and should be explored in future work.

Importantly, frequency combs are compatible with photonic integration and are rapidly being miniaturized through these approaches[64,65]. Active research in this direction has provided lasers and frequency comb generators that are now fully integrated on chip. We envision that these comb sources will be combined with on-chip amplification or small-scale fiber amplifiers to drive ultra-efficient nanophotonic nonlinear optics that can project the near-infrared spectra into the mid-infrared. Such a platform could occupy just a few liters of volume and holds the potential for significantly lower cost and complexity than the bulk laser frequency comb generator shown here. A fully integrated DCS instrument would significantly reduce the size, weight, power, and cost of each sensor and truly enable a large-scale distributed sensor network for chemical-specific aerosol monitoring.

In summary, we have presented a novel technique to perform standoff aerosol identification using mid-infrared dual-comb spectroscopy and machine-learning-based identification algorithms. Ongoing work to produce compact DCS systems will reduce the size and cost of these systems, enabling large-scale deployments for aerosol monitoring. This novel technique for standoff aerosol monitoring has broad applicability including aerosol threat monitoring for public safety and security, monitoring pollutant aerosol emissions for public health, and enabling new studies of reacting aerosols in laboratory environments.

## Methods

### *Aerosol generation and characterization*

A novel aerosol chamber was constructed to facilitate standoff spectroscopic measurements. The body of the chamber is a 1.57 m long, 30 cm diameter galvanized steel duct. Each end is capped with a 30 cm diameter galvanized steel endcap with several holes drilled into each to install O-ring-sealed Swagelok ports. All injection ports were installed in one endcap, and all sampling and exhaust ports fitted to the opposite endcap to ensure enough residence time between addition and sampling for adequate mixing. A small mixing fan (10.6 $W/m^3$ of power per unit volume of the chamber) was also installed inside the chamber which allowed the contents to reach 90% mixing within 7 seconds of injection. The fan was typically activated for 10 seconds, thus using ~106 $J/(m^3*h)$ (assuming a typical use of once per hour). $CO_2$ and $H_2O$ concentrations in the chamber were monitored with a LI-COR LI-840A gas monitor. The DCS laser was transmitted through the chamber via 5 cm diameter ZnSe windows installed on both sides. The windows were flushed with dry air to limit deposition of compounds tested in the chamber.

Three different aerosol generation methods were used, depending on the specifications of each experiment. For the generation of aerosols from water-soluble compounds, a TSI 3076 Collison atomizer was typically used. The Collison atomizer is ideal for consistent generation of high concentrations of <1 μm particles over 5–20-minute timescales. For the addition of non-water-soluble organic aerosols, two different methods were constructed and used. The first is a vaporizer based on a design developed by Finewax et al.[66]. This method is ideal for adding large amounts of aerosols very quickly (<1-minute timescales) but provides limited control over the sizes of the aerosols generated. It can also be used for solid compounds dissolved in appropriate solvents. For better control of the size distributions for organic liquids, an evaporation/condensation setup was used, based on the original design by Sinclair and La Mer[67]. This functions similarly to the vaporizer, but generates aerosols much more slowly, and with controls for changing the sizes and concentrations generated more precisely[68]. For the oleic acid particle additions discussed in this work, the evaporation/condensation apparatus was used with pure oleic acid liquid (200 °C, flowing $N_2$ gas). For the benzyl fentanyl additions, the pure compound was dissolved in pure water (0.01–0.10 $g/cm^3$) and injected with the vaporizer (250 °C, flowing $N_2$ gas). Using the vaporizer rather than the Collison atomizer allowed for rapid, discrete particle addition steps (e.g., Figures 3, 4) and minimal use of the relatively expensive analyte.

For the monitoring and characterization of aerosol concentrations in the chamber, multiple instruments were used. The first of these was the Scanning Mobility Particle Sizer (SMPS) from TSI including a 3010 condensation particle counter, 3080 classifier, and 3081 differential mobility analyzer. The SMPS measured particle sizes from 15–600 nm at a time resolution of 2 minutes. An Aerodynamic Particle Sizer (APS): TSI 3321 with a 3306 diluter (converted "impactor inlet") was used to measure particles from 500 nm – 20 μm at a time resolution of 20 seconds. Together, the SMPS and APS provide a very broad size range of particle detection, including effectively all aerosols generated in these experiments. The size distributions of aerosols generated frequently fall into the overlap region (500–600 nm) of the two instruments, allowing for direct comparison and verification.

*Dual-comb spectroscopy*

The fundamentals of dual-comb spectroscopy (DCS) and its specific application to open-path aerosol sensing are described here in greater detail for reference. DCS utilizes two optical frequency combs to perform a spectrally resolved measurement using a single photodetector without any moving parts or dispersive optics. Figure 5 illustrates the operating principle of DCS in the frequency domain (a) and the time domain (b). In the frequency domain, Figure 5(a), each frequency comb outputs a spectrum of discrete spectral elements (i.e., comb teeth), and each comb tooth is perfectly spaced by a fixed interval defined by the repetition rate of the source laser, $f_r$. The two frequency combs are operated with repetition rates differing by $\Delta f_r$. As a result, adjacent pairs of comb teeth will have slightly different optical frequencies, and the spacing between comb teeth is unique for each pair of comb teeth. When these two spectra are combined and measured on a photodetector, adjacent comb teeth (one from each comb) produce a heterodyne beat in the radio frequency (RF) domain at the difference between their optical frequencies. This process occurs for each pair of comb teeth producing an RF spectrum that perfectly recreates the optical spectrum, including any absorption signatures, with RF comb teeth spaced by $\Delta f_r$.

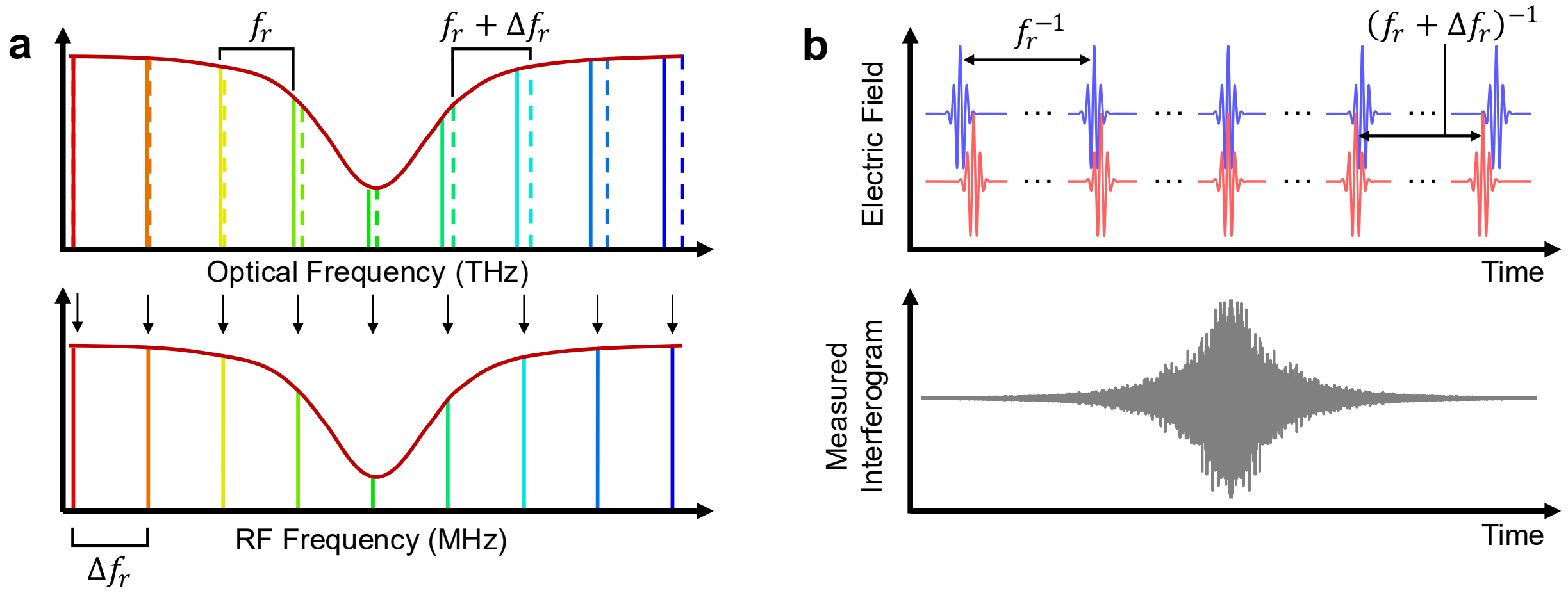

*Figure 5: Operating principle of dual-comb spectroscopy illustrated in both the frequency domain (a) and time domain (b).*

In the time domain, Figure 5(b), DCS is described by the interaction between pulses emitted by each frequency comb laser. If two laser pulses are initially overlapped in time, subsequent pulses will be slightly offset in time because the pulse rate of the lasers differs by $\Delta f_r$. The pulses will overlap again at intervals of $1/\Delta f_r$. When a pair of pulses are incident on a photodetector and the signal sampling rate is synchronized with the pulse rate of one laser, each sampled data point corresponds to the product of the electric fields of the two pulses and a single measured point of an interferogram. As the time delay between pulses varies over a duration of $1/\Delta f_r$, an entire interferogram is measured. This is directly analogous to an interferogram in standard FTS, but the optical path delay between two paths is replaced by a time delay between two laser pulses. The spectrum is given by the Fourier transform of the interferogram, and the zero-path delay point in the interferogram is defined by the time where both pulses are exactly overlapped.

When DCS is applied to open-path aerosol measurements, individual comb teeth are absorbed and scattered by gases and aerosol particles along the path, encoding information about the chemical constituents and thermodynamic conditions in the transmitted dual-comb spectrum. Gases and aerosols that are uniformly present along the path reduce the intensity of the transmitted light according to the Beer-Lambert law:

$$I_t(\nu) = I_0(\nu) \exp\left[-\left(\sum_i^{gases} \sigma_i(\nu)N_i + \sum_j^{aerosols} \sigma_j(\nu)M_j\right)L\right]. \quad (1)$$

Here, $I_t(\nu)$ is the transmitted optical intensity (i.e., irradiance in W/cm$^2$) at optical frequency $\nu$, $I_0$ is the incident optical intensity, $\sigma_i$ is the absorption cross section of gas species $i$, $N_i$ is the number density of gas species $i$, $\sigma_j$ is the ensemble-average (across particle size and morphology) extinction cross section of aerosol species $j$, which includes absorption and scattering, $M_j$ is the mass concentration of aerosol species $j$, and $L$ is the path length through the gases and aerosols. With measurements of $I_t$ and measurements or estimates of $I_0$, the extinction (or absorbance if there are no scattering particles) can be calculated as $\tau_{ext} = -ln(I_t/I_0)$. This quantity is a sum of the absorbance or extinction from all chemical species and is used here for the identification of aerosol species.

The absorption and extinction signatures of small gases and aerosols have significant differences in both wavelength dependence and magnitude at the relevant concentrations, which are illustrated in Figure 6. Figure 6(a) shows simulated absorbance from atmospheric gases ($H_2O$, $CO_2$, $CH_4$) and simulated extinction from benzyl fentanyl, and Figure 6(b) shows simulated incident and transmitted laser intensity spectra for a DCS measurement of the combined spectrum. The small gas-phase molecules have absorption spectra consisting of narrow absorption lines arising from transitions between individual rovibrational energy levels, and the peak magnitude of the absorbance is relatively large. In contrast, the extinction spectra of aerosols exhibit broad features, and the peak absorbance magnitude is 2–3 orders of magnitude weaker than the gas-phase absorption due to the relatively low aerosol concentration. The sum of these spectra is dominated by the gas-phase absorption, which must be removed from measured data to reveal the underlying aerosol signatures. In the transmitted intensity spectrum, Figure 6(b), the gas-phase signatures are clearly distinguishable from the laser's incident intensity, but the aerosol signatures appear only as a small offset compared to the incident intensity. DCS measurements sample the combined spectrum at a high spectral sampling rate defined by $f_r$ (~0.01 cm$^{-1}$) and with negligible instrument broadening defined by the linewidth of individual comb teeth (~1x10$^{-6}$ cm$^{-1}$). As a result, DCS measurements fully resolve the gas-phase absorbance, which can be subtracted from the combined extinction spectrum to produce a pure aerosol extinction spectrum.

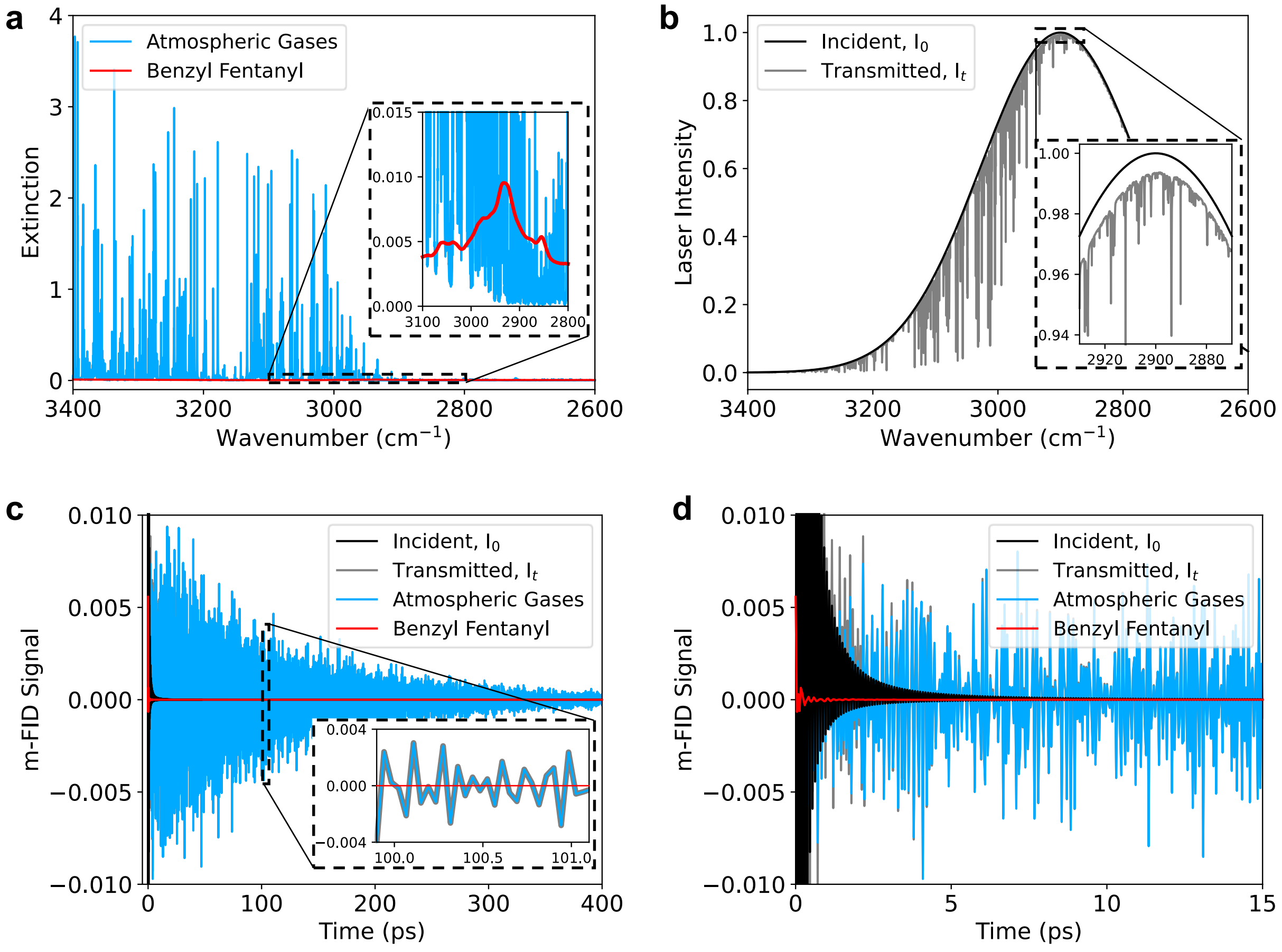

*Figure 6: (a) Simulated absorbance and extinction signals for atmospheric gases and benzyl fentanyl. (b) Simulated incident and transmitted laser intensities with an idealized Gaussian laser spectrum. (c) Time-domain m-FID signals corresponding to the transmitted intensity, incident laser spectrum, and benzyl fentanyl extinction. (d) Zoom view of the m-FID signals at early times.*

Figure 6(c-d) show time-domain representations of the absorbance, extinction, and laser spectra which are used to find the best-fit gas-phase absorbance. Following the methodology of Cole, Makowiecki et al.[54], the time-domain signals are calculated as the inverse Fourier transform of the gas-phase absorption or aerosol extinction signatures, providing the free induction decay (FID) of the molecules, and the inverse Fourier transform of the negative natural logarithm of the incident and transmitted intensities, providing the modified free induction decay (m-FID) of the laser signals. This formulation produces an additive relationship between the m-FID of the incident intensity and the FID of the gases and aerosols:

$$\mathcal{F}^{-1}\{-\ln(I_t)\} = \mathcal{F}^{-1}\{-\ln(I_0)\} + \mathcal{F}^{-1}\{\tau_{ext,aerosol}\} + \mathcal{F}^{-1}\{\alpha_{gas}\}. \quad (2)$$

Here, $\mathcal{F}^{-1}$ is the inverse Fourier transform, $\tau_{ext,aerosol}$ is the total extinction of all aerosol components and $\alpha_{gas}$ is the total absorbance of all gases. The m-FID of the incident intensity and the FID of the aerosol signatures quickly decay to zero, and the m-FID of the transmitted intensity is equal to the FID of the gas-phase signal at later times ($t > 10$ ps). As a result, a model of the gas-phase absorbance can be fit to the transmitted intensity in the time domain without knowledge of the laser's incident intensity or the aerosol signatures if only signals at later times are considered. This method is used to find the best-fit gas-phase absorbance spectrum, which is then removed from the measured transmitted intensity to produce a signal only containing the aerosol extinction and the incident laser intensity.

The challenge for aerosol analysis, compared to gas-phase analysis, arises from the similarity between the laser's incident intensity spectrum and the aerosol extinction signatures at the concentrations of interest. In the time domain, Figure 6(d), the benzyl fentanyl FID signal completely decays before the m-FID signal of the incident intensity, so the time-gated analysis used for gas-phase fitting cannot be applied for aerosol analysis. As a result, the aerosol signatures must be calculated using the measured laser spectra in the frequency domain, and a measurement or calculation of $I_0$ is required. In the frequency domain, Figure 6(b), the benzyl fentanyl signature is visible as the spectrally broad gap between the incident and transmitted intensities in the zoom panel. The benzyl fentanyl signature is similar to that of a broad decrease in $I_0$ due to a fluctuation in the laser's output spectrum or intensity. Because the aerosol absorbance signals of interest are on the order of $10^{-2}$–$10^{-3}$, the relative fluctuations in $I_0$ on the order of 0.1% can prevent aerosol identification. To address this, we utilize a reference detector to monitor the incident intensity spectrum in parallel with the aerosol measurement. The signals from each of the detectors are utilized in the pre-processing procedure to calculate the pure aerosol extinction spectrum.

*Data pre-processing*

Data pre-processing describes all calculations required to convert the raw DCS data acquired with this system to a measured aerosol extinction spectrum. This section describes all pre-processing steps before measured extinction spectra are passed to the aerosol identification algorithms. Pre-processing consists of three general steps: (1) time-averaging raw data, (2) correction for gas-phase absorption, (3) calculation of the extinction spectrum.

For this work, $\Delta f_r$ was set to approximately 1 kHz, so each individual interferogram consisted of approximately 1 ms of data. Many interferograms were averaged together in the time domain to increase the integration time of the measurement and improve the signal-to-noise ratio (SNR). To ensure constructive interference between frequency components when averaging interferograms in the time domain, the phase of each measured interferogram must be aligned. This alignment procedure, called phase correction, was implemented by shifting the phase of the frequency components in an interferogram to match those of a reference interferogram. For these measurements, a data acquisition system (DAQ) with integrated real-time phase correction was utilized to phase correct and average data together[69]. The DAQ continually acquired, phase corrected, and averaged interferograms measured from the reference and signal photodetectors. Averaged interferograms were saved to the disk at a rate of one interferogram per second, which corresponded to roughly 1000 raw interferograms averaged together in each saved file. These one-second interferograms were added together to produce a total integration time of five minutes per measurement. A Fourier transform was applied to each averaged interferogram to produce the measured intensity spectrum.

Measured spectra, typically consisting of five minutes of averaged data, were then corrected to remove background gas absorption. A gas-phase absorbance spectrum including contributions from $H_2O$, $CO_2$, and $CH_4$ was simulated using the HITRAN2020 database[70] and fit to each measured spectrum by varying the gas composition and temperature following the Levenberg-Marquardt algorithm. The spectra were fit in the time domain using the m-FID technique developed by Cole, Makowiecki et al.[54], which reduces the influence of laser baseline structure on spectral fitting. The high spectral resolution of the DCS system (0.033 $cm^{-1}$) clearly resolved the gas-phase absorption signatures, enabling a high-fidelity fit to the gas-phase signatures. The gas-phase absorption signatures were then removed from the measured transmitted intensity spectrum using the following equation:

$$I_{corrected}(\nu) = I_{meas}(\nu)\exp(\alpha_{fit}(\nu)). \quad (3)$$

Here, $I_{corrected}(\nu)$ is the corrected spectrum with the gas-phase absorption removed, $I_{meas}(\nu)$ is the measured transmission spectrum, and $\alpha_{fit}(\nu)$ is the best-fit absorbance spectrum.

For each aerosol test, an aerosol extinction spectrum was calculated using four corrected transmission spectra. The four spectra were measurements from both the signal and reference detectors for both a background measurement without aerosols present and the test measurement with aerosols present. The aerosol extinction spectrum was calculated using the following equation:

$$\tau_{aerosol}(\nu) = -\ln\left[\frac{I_{sig}(\nu)}{I_{ref}(\nu)}\right]_{test} + \ln\left[\frac{I_{sig}(\nu)}{I_{ref}(\nu)}\right]_{bg}. \quad (4)$$

Here, $\tau_{aerosol}(\nu)$ is the aerosol extinction at optical frequency $\nu$ and $I$ is the corrected transmitted intensity. The subscripts *sig* and *ref* refer to the signal and reference detectors, respectively, and the subscripts *test* and *bg* refer to the test and background measurements, respectively. In standard absorption spectroscopy terms, the signal data is the transmitted intensity, and the reference data is the incident intensity. Ideally, the first term in Equation (4) would produce the aerosol extinction spectrum. However, wavelength-dependent variations in the transmittance and reflectance of transceiver optics result in differences between the measured signal- and reference-branch spectra. The background term, measured without an aerosol sample present, is needed to correct for these differences. In practice, the spectral shape of this background term exhibited a slow drift over time. To reduce the influence of this background drift, two background spectra were measured: one before the aerosol addition and one after the chamber was evacuated. A final background spectrum was calculated by interpolating between these two measured background spectra based on the acquisition time of the test measurement. The corrected intensity spectra typically contained optical interference fringes (i.e., etalons), which are clearly visible in the zoom panel of Figure 2. These etalons largely cancelled out during the normalization applied with Equation (4), and the etalons remaining in the aerosol extinction spectrum were removed by filtering out their signatures in the time-domain representation of the aerosol extinction spectrum[54].

*Aerosol identification algorithms*

The general approach is to use a database of synthetic extinction spectra spanning various species and particle sizes for chemical identification from measured spectra. The synthetic extinction spectra were generated from complex refractive indices of organic compounds and a radiative transfer model. The identification algorithm differs from typical approaches for single molecule identification in homogeneous samples in that species are identified in binary mixtures in large particles with nonnegligible scattering contributions.

<u>Derivation of complex refractive index of benzyl fentanyl</u>

Due to the lack of available complex refractive index data for the key species of our study, benzyl fentanyl, an empirical approach was adopted to derive the real ($n$) and imaginary ($k$) parts of its refractive index. The absorbances of benzyl fentanyl dissolved in acetonitrile (ACN) and in powder form were measured by ATR-FTIR and were used to derive the $k$ vectors. To account for the baseline drift of the instrument, the mean absorbance between 3950 and 4000 $cm^{-1}$ was subtracted from the measured absorbance spectra to keep the value of k near zero at the lower end of wavelength (4000 $cm^{-1}$), where no light absorption was expected. Subsequently, the $n$ vector was estimated over the IR region via a subtractive Kramers-Kronig (KK) transformation. This method of $n$ vector derivation requires an anchor refractive index ($n_{\text{anchor}}$) at a specific wavenumber, where the value of k is expected to be near zero. Two independent methods of estimating the value of $n$ at the visible region (i.e., 589 nm) showed a consistent value of $n$ (1.57 by the graph machine model by Duprat et al.[71] and 1.57–1.65 based on elemental ratios by Li et al.[72]). Since the Sellmeier equation[73] indicates a negligible refractive index difference (≈ 0.02) between 589 nm and 2500 nm (4000 cm-1), the value of $n_{\text{anchor}}$ at 4000 $cm^{-1}$ was approximated to be 1.55 (i.e., 1.57 minus 0.02).

The derived optical constants were subsequently utilized to simulate aerosol extinction and derivative spectra for log-normally distributed benzyl fentanyl particles across varying geometric mean diameters of 0.3, 1, 4, 7, and 10 μm. To systematically evaluate the impact of the assumed value of $n_{\text{anchor}}$ on simulated aerosol spectra, we compared the simulated aerosol spectra with $n_{\text{anchor}}$ values ranging from 1.45 to 1.65. The results indicated that for sub-micron particles (0.3 and 1 μm), varying the value of $n_{\text{anchor}}$ had negligible impact on both the extinction and derivative spectra. For larger particles (4 to 10 μm), the choice of $n_{\text{anchor}}$ caused noticeable baseline shifts and magnitude differences in the extinction spectra. However, the resulting derivative spectra remained highly robust and virtually identical across the tested values of $n_{\text{anchor}}$, regardless of particle size, proving the robustness of the derived complex refractive index of benzyl fentanyl for the use of its detection.

Simulation of synthetic aerosol spectra

We selected approximately 100 target chemicals spanning pharmaceutical agents, plasticizers, organophosphorus, nitro organics, and surrogate compounds that are related to defense and/or industrial sensing studies[74]. For each target chemical, we compiled its complex refractive index ($m = n + ki$) from PNNL databases[53] interpolated to a uniform wavenumber grid of 2 $cm^{-1}$ for a wavenumber region from 800 to 4000 $cm^{-1}$. We simulated extinction, forward-scattering, and backward-scattering efficiencies ($Q_{ext}$, $Q_{fs}$, and $Q_{bs}$) via standard Lorenz–Mie theory using the custom MATLAB functions[75]. Their cross sections ($CS_{ext}$, $CS_{fs}$, and $CS_{bs}$) were calculated as follows:

$$CS_{ext/fs/bs}(\tilde{\nu}, D_p) = \frac{\pi {D_p}^2}{4} Q_{ext/fs/bs}(\tilde{\nu}, D_P, m)$$

where $\tilde{\nu}$ is wavenumber and $D_p$ is a particle diameter.

We followed the liquid aerosol scattering model[76] to construct synthetic IR spectra of the target chemicals. This model simulated a condition where an incident laser beam from an instrument went through an aerosol plume, reflected at a surface of a given reflective characteristic, went through the same aerosol plume again, and was collected by a detector situated at the incident beam location. The following equation therefore simulated the fraction of light reaching back to the detector, which was then $log_{10}$-transformed to produce an extinction spectrum:

$$\begin{aligned} I_{det}(\tilde{\nu}) = \Bigg[ & I_0(\tilde{\nu}) \left( \frac{\sum CS_{bs}(\tilde{\nu})}{A} + \rho(\tilde{\nu}) \frac{\sum CS_{fs}(\tilde{\nu})}{A} \right) \frac{\Omega}{4\pi} \\ & + I_0(\tilde{\nu}) \rho(\tilde{\nu}) \left( 1 - \frac{\sum CS_{ext}(\tilde{\nu})}{A} \right) \left( \rho(\tilde{\nu}) \frac{\sum CS_{bs}(\tilde{\nu})}{A} + \frac{\sum CS_{fs}(\tilde{\nu})}{A} \right) \frac{\Omega}{4\pi} \\ & + I_0(\tilde{\nu}) \rho(\tilde{\nu}) \left( 1 - \frac{\sum CS_{ext}(\tilde{\nu})}{A} \right)^2 \Bigg] \end{aligned}$$

where $I_{det}$ is an intensity of the laser beam detected at the detector at a given wavenumber, $I_0$ is an initial intensity of the laser beam at a given wavenumber, $\Sigma CS_{ext/fs/bs}$ is the total extinction/forward-scattering/backward-scattering cross sections integrated over a particle size of 0.05–15 μm at a given wavenumber, A is the cross-sectional area of the laser beam, ρ is the reflectance of the surface at a given wavenumber, and Ω is the solid angle subtended by the detector.

For each target chemical, we simulated 1000 spectra with varying log-normal aerosol size distributions (mean diameter ranging from 0.1 to 10 μm with a standard deviation of 0.1 to 0.5 multiplied by the given mean diameter) and concentration (1 to 10 $mg/m^3$). Unless otherwise noted, we assumed single-component aerosol, and for aerosol mixtures, we assumed chemicals were externally mixed (i.e., extinction spectra were linearly added).

Chemical identification algorithm

Before the measured data were fed into the classifier, the spectra underwent further processing steps, as outlined in Figure 2(a). Min-max scaling on extinction spectra was performed to eliminate potential impacts of linear baseline drift, though this process also removed the information about aerosol concentration. We used a conventional Savitzky–Golay filter to smooth extinction spectra. Smoothness is critical to obtain robust derivative spectra, which were obtained via second-order numerical differentiation. Derivative spectra were then standardized to have a mean of 0 and a standard deviation of 1. To further remove noise and to focus on important features, we performed principal component analysis (PCA) and extracted low-dimensional projections (reducing the number of features approximately by an order of magnitude) to use as features.

Identification of unknown chemical(s) from a measured aerosol extinction spectrum was performed via a supervised machine learning algorithm, namely an artificial neural network (NN). A feedforward NN consists of an input layer that accepts the low-dimensional projections by the PCA, several fully connected hidden layers, and an output layer. Each neuron in a hidden layer computes a weighted sum of the outputs from the previous layer and passes this result through a nonlinear rectified linear unit (ReLU) activation function. The output layer employs a softmax activation function, which converts the output scores into a probability distribution over the entire set of possible classes that represent the identity of ~100 chemicals included in the training dataset.

The NN-based multi-class classification model described above predicts the identity of chemicals from the pool of known chemicals included in the training dataset. Therefore, the model will inevitably encounter misclassification when the input is of unknown chemicals that are not included in the training dataset. To avoid misclassification in case the input is foreign to the trained model, we introduced an additional layer in the algorithm to filter out unknown chemicals via anomaly detection. We used a distance metric to verify if the predicted chemical is one of the chemicals in the training dataset. The normalized Mahalanobis distance between the input extinction spectrum and the simulated spectrum of the predicted chemical was calculated as follows:

$$D_{AD} = \frac{\sqrt{(X_{input} - X_{sim})S^{-1}(X_{input} - X_{sim})}}{\sqrt{X_{sim}S^{-1}X_{sim}}}$$

where $X_{\mathrm{input}}$ is an input feature, $X_{\mathrm{sim}}$ is a low-dimensional projection based on the simulated spectrum of the predicted chemical, and S is a covariance matrix. If the calculated distance value is low, it means that the input spectrum is similar to a simulated spectrum of the predicted chemical. Since the equation provides a normalized distance, we set the threshold value of 1 to determine whether to confirm or to negate the chemical identification result of the model.

Identification of chemicals in mixtures

The above chemical identification algorithm is primarily designed to determine the identity of an individual chemical included in the training dataset. We also applied this algorithm to binary aerosol mixtures. During these measurements of aerosol mixtures, one chemical was introduced first, followed by a second chemical. To decouple and identify the presence of both chemicals, we employed the two following approaches.

The first approach was where we assumed the availability of reliable, continuous measurements of the background spectra. Since the chemicals were added in a sequential manner, the background spectrum that represents the first chemical was subtracted from the next measured spectrum that represents chemical mixtures. We then utilized the classifier to identify the first chemical directly from the background spectrum and the second chemical from the background-subtracted spectrum. The second approach was used to identify the two chemicals in a sequential manner without relying on prior background knowledge or measurement. This approach relies on the assumption that the dominant chemical exhibits the most prominent spectral features. The model was used to initially identify the primary chemical from the spectrum that represents mixtures. Once the primary chemical was identified, the most similar simulation spectrum of the identified chemical was used to obtain a residual spectrum via asymmetric least squares approach such that the residual spectrum consists of positive values. The model was run using this residual spectrum to identify the second chemical.

*Outdoor Extended Standoff Distance Measurement*

A fieldable mid-infrared dual-comb spectrometer was used to identify aerosols at a 100 m standoff distance outdoors. The spectrometer has a 200 MHz repetition rate, uses intra-pulse difference frequency generation to convert 1550 nm light to the mid-infrared, and was engineered into a rack-mount for field use. A retroreflector was placed outdoors 100 m away from the laser transceiver, and the laser passed through uncontrolled outdoor air approximately 1 m above the ground. A dibutyl phthalate sample was prepared and deposited between two 5 cm diameter potassium bromide optical windows with an approximate aerosol loading of 450 $mg/m^2$ [77], and the sample was placed in the beam path for 10 minutes while data was collected and averaged. Figure 7(a) shows the experimental setup schematically, and images of the setup are in Figure 7(b). The measured data and first derivative spectra are in Figure 7(c) and Figure 7(d), respectively. Spikes at higher wavenumbers in the raw measurement in Figure 7(c) are residuals from water vapor signature removal during pre-processing. The long path length and humid outdoor conditions caused water vapor absorption features to saturate the laser transmission. Fitting near saturated features is sensitive to small imperfections in the absorption model or handling of the laser baseline and can lead to significant residuals. Nevertheless, the bin averaging and neural network successfully identified dibutyl phthalate in these conditions.

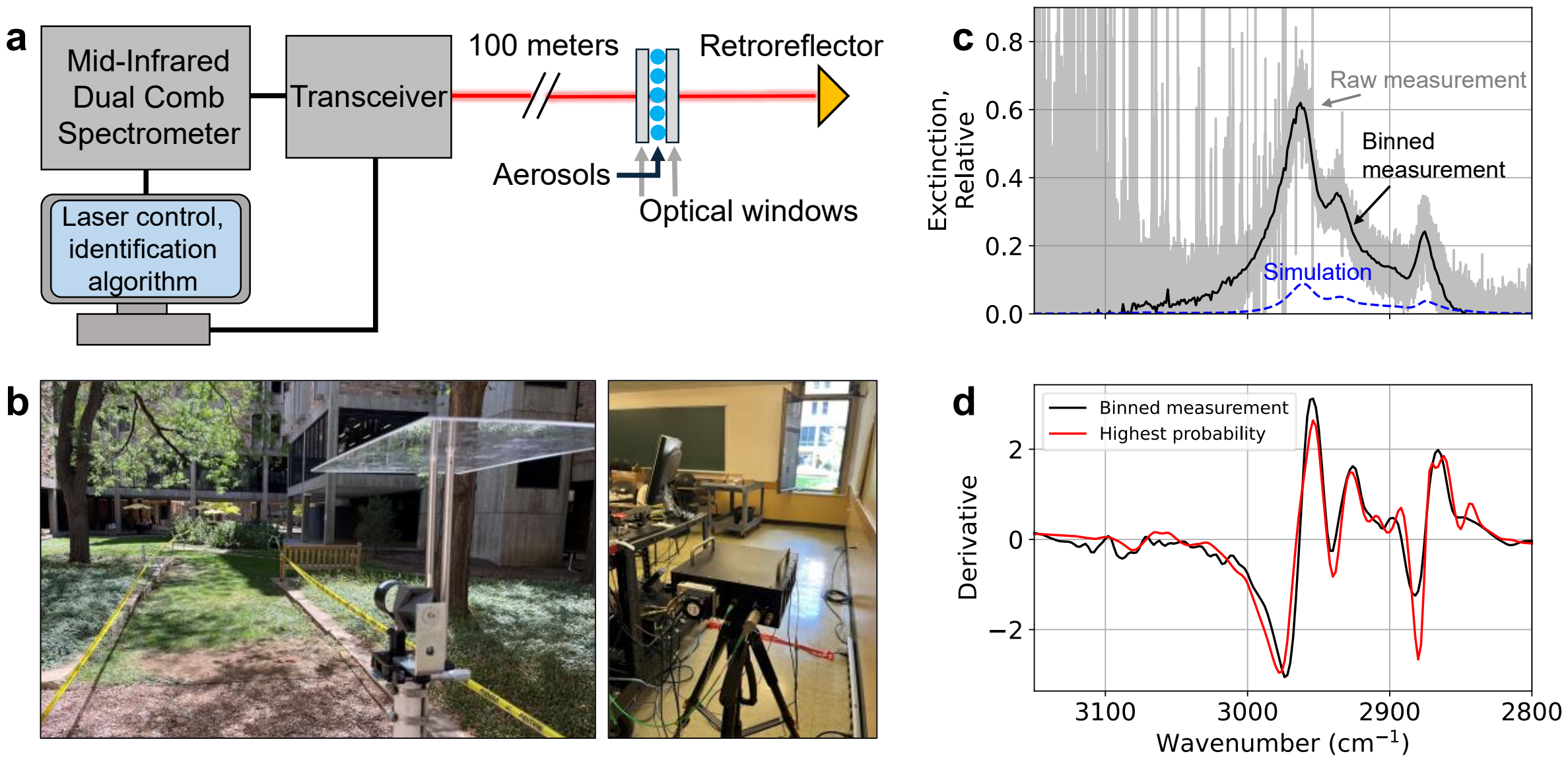

*Figure 7: (a) Experimental setup for outdoor extended standoff distance measurement. The aerosols are contained between two optical windows and placed in the 100 m beam path. (b) Aerosol sample and retroreflector set up with long optical path (left), and laser transceiver pointing out the window (right). (c) The relative extinction spectrum in raw and binned forms with comparison to simulated dibutyl phthalate absorbance. The sharp spikes in the raw measurement are residuals from saturated water signatures. (d) The derivative spectrum of the binned measurement and the highest-probability derivative spectrum returned from the neural network.*

## References

1. Wattana, M. & Bey, T. Mustard gas or sulfur mustard: an old chemical agent as a new terrorist threat. *Prehosp. Disaster Med.* **24**, 19–29 (2009).
2. Ganesan, K., Raza, S. K. & Vijayaraghavan, R. Chemical warfare agents. *J. Pharm. Bioallied Sci.* **2**, 166–178 (2010).
3. Nepovimova, E. & Kuca, K. Chemical warfare agent NOVICHOK – mini-review of available data. *Food Chem. Toxicol.* **121**, 343–350 (2018).
4. Riches, J. R., Read, R. W., Black, R. M., Cooper, N. J. & Timperley, C. M. Analysis of clothing and urine from Moscow theatre siege casualties reveals carfentanil and remifentanil use. *J. Anal. Toxicol.* **36**, 647–656 (2012).
5. Heslop, D. J. & Blain, P. G. Threat potential of pharmaceutical based agents. *Intell. Natl Secur.* **35**, 539–555 (2020).

6. Dockery, D. W. *et al.* An association between air pollution and mortality in six U.S. cities. *N. Engl. J. Med.* **329**, 1753–1759 (1993).

7. Pope III, C. A. What do epidemiologic findings tell us about health effects of environmental aerosols? *J. Aerosol Med.* **13**, 335–354 (2000).

8. Lighty, J. S., Veranth, J. M. & Sarofim, A. F. Combustion aerosols: factors governing their size and composition and implications to human health. *J. Air Waste Manag. Assoc.* **50**, 1565–1618 (2000).

9. Arfin, T. *et al.* An overview of atmospheric aerosol and their effects on human health. *Environ. Sci. Pollut. Res.* **30**, 125347–125369 (2023).

10. Zhang, Q. *et al.* Understanding atmospheric organic aerosols via factor analysis of aerosol mass spectrometry: a review. *Anal. Bioanal. Chem.* **401**, 3045–3067 (2011).

11. DeCarlo, P. F. *et al.* Field-deployable, high-resolution, time-of-flight aerosol mass spectrometer. *Anal. Chem.* **78**, 8281–8289 (2006).

12. Canagaratna, M. R. *et al.* Chemical and microphysical characterization of ambient aerosols with the Aerodyne aerosol mass spectrometer. *Mass Spectrom. Rev.* **26**, 185–222 (2007).

13. Laskin, A., Laskin, J. & Nizkorodov, S. A. Mass spectrometric approaches for chemical characterisation of atmospheric aerosols: critical review of the most recent advances. *Environ. Chem.* **9**, 163–189 (2012).

14. Maria, S. F., Russell, L. M., Turpin, B. J. & Porcja, R. J. FTIR measurements of functional groups and organic mass in aerosol samples over the Caribbean. *Atmos. Environ.* **36**, 5185–5196 (2002).

15. Liu, S., Takahama, S., Russell, L. M., Gilardoni, S. & Baumgardner, D. Oxygenated organic functional groups and their sources in single and submicron organic particles in MILAGRO 2006 campaign. *Atmos. Chem. Phys.* **9**, 6849–6863 (2009).
16. Takahama, S. *et al.* Organic functional groups in aerosol particles from burning and non-burning forest emissions at a high-elevation mountain site. *Atmos. Chem. Phys.* **11**, 6367–6386 (2011).
17. Weakley, A. T., Takahama, S. & Dillner, A. M. Ambient aerosol composition by infrared spectroscopy and partial least-squares in the chemical speciation network: organic carbon with functional group identification. *Aerosol Sci. Technol.* **50**, 1096–1114 (2016).
18. Sze, S.-K., Siddique, N., Sloan, J. J. & Escribano, R. Raman spectroscopic characterization of carbonaceous aerosols. *Atmos. Environ.* **35**, 561–568 (2001).
19. Craig, R. L., Bondy, A. L. & Ault, A. P. Surface enhanced Raman spectroscopy enables observations of previously undetectable secondary organic aerosol components at the individual particle level. *Anal. Chem.* **87**, 7510–7514 (2015).
20. Marina-Montes, C. *et al.* Characterization of atmospheric aerosols in the Antarctic region using Raman spectroscopy and scanning electron microscopy. *Spectrochim. Acta A Mol. Biomol. Spectrosc.* **266**, 120452 (2022).
21. Estefany, C., Sun, Z., Hong, Z. & Du, J. Raman spectroscopy for profiling physical and chemical properties of atmospheric aerosol particles: a review. *Ecotoxicol. Environ. Saf.* **249**, 114405 (2023).
22. Comerón, A., Muñoz-Porcar, C., Rocadenbosch, F., Rodríguez-Gómez, A. & Sicard, M. Current research in lidar technology used for the remote sensing of atmospheric aerosols. *Sensors* **17**, 1450 (2017).

23. Burton, S. P. *et al.* Aerosol classification using airborne high spectral resolution lidar measurements – methodology and examples. *Atmos. Meas. Tech.* **5**, 73–98 (2012).

24. Burton, S. P., Vaughan, M. A., Ferrare, R. A. & Hostetler, C. A. Separating mixtures of aerosol types in airborne high spectral resolution lidar data. *Atmos. Meas. Tech.* **7**, 419–436 (2014).

25. Richardson, S. C. *et al.* Bioaerosol detection over Athens, Greece using the laser induced fluorescence technique. *Sci. Total Environ.* **696**, 133906 (2019).

26. Hara, Y. *et al.* Retrieval of aerosol components using multi-wavelength Mie-Raman lidar and comparison with ground aerosol sampling. *Remote Sens.* **10**, 937 (2018).

27. Arnott, W. P., Schmitt, C., Liu, Y. & Hallett, J. Droplet size spectra and water-vapor concentration of laboratory water clouds: inversion of Fourier transform infrared (500–5000 $cm^{-1}$) optical-depth measurement. *Appl. Opt.* **36**, 5205–5216 (1997).

28. Wu, C.-F., Chen, Y.-L., Chen, C.-C., Yang, T.-T. & Chang, P.-E. Applying open-path Fourier transform infrared spectroscopy for measuring aerosols. *J. Environ. Sci. Health A* **42**, 1131–1140 (2007).

29. Kira, O., Linker, R. & Dubowski, Y. Detection and quantification of water-based aerosols using active open-path FTIR. *Sci. Rep.* **6**, 25110 (2016).

30. Smekens, J.-F. *et al.* Quantification of gas, ash, and sulphate aerosols in volcanic plumes from open path Fourier transform infrared (OP-FTIR) emission measurements at Stromboli volcano, Italy. *Front. Earth Sci.* **10**, 1005738 (2023).

31. Salcido, J. M. O. *et al.* Quantitative infrared spectroscopy of aerosols: Mie theory modeling with experimental validation. *Optica* **12**, 1462–1468 (2025).

32. Maidment, L., Zhang, Z., Howle, C. R. & Reid, D. T. Stand-off identification of aerosols using mid-infrared backscattering Fourier-transform spectroscopy. *Opt. Lett.* **41**, 2266–2269 (2016).

33. Maidment, L. *et al.* Systematic spectral shifts in the mid-infrared spectroscopy of aerosols. *Opt. Express* **26**, 18975–18981 (2018).

34. Lecours, M. J. *et al.* Atlas of ACE spectra of clouds and aerosols. *J. Quant. Spectrosc. Radiat. Transf.* **292**, 108361 (2022).

35. Gurton, K. P., Ligon, D. & Dahmani, R. Measured infrared optical cross sections for a variety of chemical and biological aerosol simulants. *Appl. Opt.* **43**, 4564–4570 (2004).

36. Schiller, S. Spectrometry with frequency combs. *Opt. Lett.* **27**, 766–768 (2002).

37. Keilmann, F., Gohle, C. & Holzwarth, R. Time-domain mid-infrared frequency-comb spectrometer. *Opt. Lett.* **29**, 1542–1544 (2004).

38. Coddington, I., Newbury, N. & Swann, W. Dual-comb spectroscopy. *Optica* **3**, 414–426 (2016).

39. Rieker, G. B. *et al.* Frequency-comb-based remote sensing of greenhouse gases over kilometer air paths. *Optica* **1**, 290–298 (2014).

40. Ycas, G. *et al.* Mid-infrared dual-comb spectroscopy of volatile organic compounds across long open-air paths. *Optica* **6**, 165–168 (2019).

41. Han, J.-J. *et al.* Dual-comb spectroscopy over a 100 km open-air path. *Nat. Photon.* **18**, 1195–1202 (2024).

42. Chen, X. *et al.* Phase-sensitive open-path dual-comb spectroscopy with free-running combs. *Phys. Rev. Appl.* **19**, 044016 (2023).

43. Timmers, H. *et al.* Molecular fingerprinting with bright, broadband infrared frequency combs. *Optica* **5**, 727–732 (2018).

44. Makowiecki, A. *et al.* Mid-infrared dual frequency comb spectroscopy for combustion analysis from 2.8 to 5 microns. *Proc. Combust. Inst.* **38**, 1627–1635 (2020).

45. Luo, D. *et al.* Parallel sensing of multiple greenhouse gases adopting a mid-infrared dual-comb spectrometer with 300,000 comb-tooth-resolved frequency components. *Photon. Res.* **13**, A1–A7 (2025).

46. Konnov, D., Muraviev, A., Vasilyev, S. & Vodopyanov, K. High-resolution frequency-comb spectroscopy with electro-optic sampling and instantaneous octave-wide coverage across mid-IR to THz at a video rate. *APL Photonics* **8**, 110801 (2023).

47. Hoghooghi, N. *et al.* Broadband 1-GHz mid-infrared frequency comb. *Light Sci. Appl.* **11**, 264 (2022).

48. Bohren, C. F. & Huffman, D. R. *Absorption and Scattering of Light by Small Particles*. (John Wiley & Sons, 2008).

49. Xu, M., Wang, C.-H., Terracciano, A. C., Masunov, A. E. & Vasu, S. S. High accuracy machine learning identification of fentanyl-relevant molecular compound classification via constituent functional group analysis. *Sci. Rep.* **10**, 13569 (2020).

50. Wang, T., Tan, Y., Chen, Y. Z. & Tan, C. Infrared spectral analysis for prediction of functional groups based on feature-aggregated deep learning. *J. Chem. Inf. Model.* **63**, 4615–4622 (2023).

51. Jung, G., Jung, S. G. & Cole, J. M. Automatic materials characterization from infrared spectra using convolutional neural networks. *Chem. Sci.* **14**, 3600–3609 (2023).

52. Kim, H. *et al.* Calculation of the infrared spectra of organophosphorus compounds and prediction of new types of nerve agents. *New J. Chem.* **46**, 8653–8661 (2022).

53. Lockwood, S. P. *et al.* Modeling aerosol transmission spectra from n(λ) and k(λ) infrared optical constants measurements of organic liquids and solids. *Opt. Express* **32**, 30169–30181 (2024).

54. Cole, R. K., Makowiecki, A. S., Hoghooghi, N. & Rieker, G. B. Baseline-free quantitative absorption spectroscopy based on cepstral analysis. *Opt. Express* **27**, 37920–37939 (2019).

55. Schauer, J. J., Kleeman, M. J., Cass, G. R. & Simoneit, B. R. T. Measurement of emissions from air pollution sources. 1. C1 through C29 organic compounds from meat charbroiling. *Environ. Sci. Technol.* **33**, 1566–1577 (1999).

56. Weitkamp, E. A. *et al.* Fine particle emission profile for a large coke production facility based on highly time-resolved fence line measurements. *Atmos. Environ.* **39**, 6719–6733 (2005).

57. Surdu, M. *et al.* Quantifying submicrometer atmospheric aerosol chemical composition using nanoelectromechanical Fourier transform infrared spectroscopy. *Sci. Adv.* **12**, eaeb2254 (2026).

58. Miller, F. J., Asgharian, B., Schroeter, J. D. & Price, O. Improvements and additions to the Multiple Path Particle Dosimetry model. *J. Aerosol Sci.* **99**, 14–26 (2016).

59. Chen, Z., Hänsch, T. W. & Picqué, N. Mid-infrared feed-forward dual-comb spectroscopy. *Proc. Natl Acad. Sci.* **116**, 3454–3459 (2019).

60. Lind, A. J. *et al.* Mid-infrared frequency comb generation and spectroscopy with few-cycle pulses and $\chi^{(2)}$ nonlinear optics. *Phys. Rev. Lett.* **124**, 133904 (2020).

61. Liu, M. *et al.* Mid-infrared cross-comb spectroscopy. *Nat. Commun.* **14**, 1044 (2023).

62. Caldwell, E. D., Sinclair, L. C., Newbury, N. R. & Deschenes, J.-D. The time-programmable frequency comb and its use in quantum-limited ranging. *Nature* **610**, 667–673 (2022).

63. Bernhardt, B. *et al.* Cavity-enhanced dual-comb spectroscopy. *Nat. Photon.* **4**, 55–57 (2010).

64. Chang, L., Liu, S. & Bowers, J. E. Integrated optical frequency comb technologies. *Nat. Photon.* **16**, 95–108 (2022).

65. Yang, Q.-F., Hu, Y., Torres-Company, V. & Vahala, K. Efficient microresonator frequency combs. *eLight* **4**, 18 (2024).

66. Finewax, Z., Jimenez, J. L. & Ziemann, P. J. Development and application of a low-cost vaporizer for rapid, quantitative, in situ addition of organic gases and particles to an environmental chamber. *Aerosol Sci. Technol.* **54**, 1567–1578 (2020).

67. Sinclair, D. & La Mer, V. K. Light scattering as a measure of particle size in aerosols. The production of monodisperse aerosols. *Chem. Rev.* **44**, 245–267 (1949).

68. Kim, D. *et al.* Development and characterization of an aircraft inlet system for broader quantitative particle sampling at higher altitudes: aerodynamic lenses, beam and vaporizer diagnostics, and pressure-controlled inlets. *Aerosol Res.* **3**, 371–404 (2025).

69. Walsh, M., Kasic, J., Cossel, K. & Genest, J. Graphics card-based real-time processing for dual comb interferometry. *Rev. Sci. Instrum.* **95**, 103005 (2024).

70. Gordon, I. E. *et al.* The HITRAN2020 molecular spectroscopic database. *J. Quant. Spectrosc. Radiat. Transf.* **277**, 107949 (2022).

71. Duprat, F., Ploix, J.-L., Aubry, J.-M. & Gaudin, T. Fast and accurate prediction of refractive index of organic liquids with graph machines. *Molecules* **28**, 6805 (2023).

72. Li, Y. *et al.* Predicting real refractive index of organic aerosols from elemental composition. *Geophys. Res. Lett.* **50**, e2023GL103446 (2023).

73. Chang, H.-J. *et al.* Refractive index measurements of liquids from 0.5 to 2 μm using Rayleigh interferometry. *Opt. Mater. Express* **14**, 1253–1267 (2024).

74. Webb, S., Coulon, F. & Temple, T. A critical review of liquid, low toxicity chemical warfare agent simulants: enhancing accuracy, safety, and methodological approaches for sampling. *J. Hazard. Mater.* **492**, 138021 (2025).

75. Mätzler, C. MATLAB functions for Mie scattering and absorption, Version 1. https://boris-portal.unibe.ch/bitstreams/afa4b3b1-5e2e-47bb-9292-963dbeec9395/download (2002).

76. Furstenberg, R., Shabaev, A., Huffman, T. J., Kendziora, C. A. & McGill, R. A. Modeling of infrared scattering signatures of liquid and solid aerosol clouds. in *Chemical, Biological, Radiological, Nuclear, and Explosives (CBRNE) Sensing XXV* vol. 13056 264 (SPIE, 2024).

77. McGill, R. A. *et al.* Static aerosol test panels as standoff detection performance targets. in *Chemical, Biological, Radiological, Nuclear, and Explosives (CBRNE) Sensing XXVII* vol. 14048 78 (SPIE, 2026).

**Funding**

Supported by the Intelligence Advanced Research Projects Activity (IARPA) via Department of Interior/ Interior Business Center (DOI/IBC) contract number 140D0423C0112. The U.S. Government is authorized to reproduce and distribute reprints for Governmental purposes notwithstanding any copyright annotation thereon. Disclaimer: The views and conclusions contained herein are those of the authors and should not be interpreted as necessarily representing the official policies or endorsements, either expressed or implied, of IARPA, DOI/IBC, or the U.S. Government.

**Acknowledgements**

The authors gratefully acknowledge the IARPA PICARD government test and evaluation team.

**Author contributions**

G.C.M. performed experiments, developed the laser data processing methodology, analyzed laser data, and managed the project. A.L. performed experiments and analyzed laser data. A.C.Z. developed the aerosol chamber, performed experiments, and analyzed aerosol data. M.T.

developed the aerosol identification algorithm and analyzed laser data. L.C.B. performed experiments and analyzed aerosol data. D.A.D. developed the aerosol chamber, performed experiments, and analyzed aerosol data. A.V.H. performed experiments and analyzed aerosol data. A.M. developed the laser data processing methodology and analyzed laser data. K.W. developed the laser transceiver and performed experiments. N.H., S.E., P.C., T.-H.W., S.A.D., G.B.R. developed the frequency comb laser source. A.H. developed the laser transceiver. G.C.M., A.L., M.T., L.C.B., D.A.D., S.T., and G.B.R. wrote the manuscript with input and review from all authors. G.B.R, G.C.M. K.V., S.A.D., J.L.J., D.H., S.T., J.B., T.K., J.L., and R.W. conceived the program, acquired funding, and supervised the research.

## Competing interest declaration

The authors declare the following competing interests: G.B.R. is a co-founder and CTO of and holds equity in LongPath Technologies. R.W. is a co-founder and chief engineer of and holds equity in LongPath Technologies. A.M., K.W., and A.H. are employees of and hold equity in LongPath Technologies. At the time of this work, T.K. was CEO, co-founder, and a board member of, and held equity in, Nexus Photonics, which was acquired by IonQ. T.K. holds equity in IonQ. At the time of this work, J.B. was chairman, co-founder, and a board member of, and held equity in, Nexus Photonics, which was acquired by IonQ. J.B. holds equity in IonQ. J.L. is president and co-founder of and holds equity in hQphotonics. K.V. is a co-founder of and holds equity in hQphotonics. T.-H.W. is founder and CEO of and holds equity in Chi3 Optics. S.A.D. holds equity in Chi3 Optics. S.T. is a co-founder and scientific advisor of and holds equity in Aerospec SA. The remaining authors declare no competing interests.

## Materials and correspondence

Please direct all correspondence and material requests to Garrett Mathews (garrett.mathews@colorado.edu)